# An Enthalpy-Based Unified Lattice Boltzmann Flux Solver for Liquid Solidification

Jinxiang Zhou[1], Liming Yang[1, 2, 3, *], Yaping Wang[1], Jie Wu[1], Xiaodong Niu[4]

[1]Department of Aerodynamics, College of Aerospace Engineering, Nanjing University of Aeronautics and Astronautics, Nanjing 210016, China

[2]State Key Laboratory of Mechanics and Control for Aerospace Structures, Nanjing University of Aeronautics and Astronautics, Nanjing 210016, China

[3]MIIT Key Laboratory of Unsteady Aerodynamics and Flow Control, Nanjing University of Aeronautics and Astronautics, Nanjing 210016, China

[4]College of Engineering, Shantou University, 243 Daxue Road, Shantou 515063, Guangdong, China

**Abstract**

An enthalpy-based uniform lattice Boltzmann flux solver (EULBFS) is proposed in this paper for simulating liquid solidification, incorporating the effects of volume expansion and shrinkage caused by density differences between liquid and solid phases. The proposed solver first establishes the relationships between the macroscopic governing equations and mesoscopic formal equations that describe the temperature, flow, and phase fields. The macroscopic governing equations are then discretized by the finite volume method (FVM), with the corresponding fluxes calculated based on the established relationships. In this way, it enables a unified and coherent solution framework for all fields. In contrast to the conventional lattice Boltzmann methods, the present approach handles additional terms directly via finite volume discretization, offering a more straightforward and flexible formulation. Furthermore, the use of the total enthalpy equation to couple the temperature field with the phase fraction allows for efficient modeling of phase change processes, significantly reducing the computational complexity associated with interface tracking. The accuracy and robustness of the proposed solver are demonstrated by a series of benchmark tests, including the conductive freezing problem, the three-phase Stefan problem, the freezing of a liquid film in a two-dimensional container, the solidification of a static droplet on a cold surface, and the freezing of a droplet upon impact with a cold surface.

---

*Corresponding author, E-mail: lmyang@nuaa.edu.cn.

1. **Introduction**

    The interfacial interaction between supercooled liquids and cooled surfaces poses significant challenges across a wide range of engineering systems and infrastructure applications. Prominent manifestations include ice accretion on aircraft engine components, which severely compromises aerodynamic performance degradation[1]-[3]; hydrodynamic profile disruption in marine propellers[4]; and catastrophic structural failures in overhead transmission lines[5]. These widespread issues have driven sustained scientific efforts to elucidate the underlying mechanisms of ice nucleation kinetics and to develop advanced anti-icing strategies. Consequently, substantial research has been dedicated to understanding the phase transition dynamics at solid-liquid interfaces and to designing surface architectures capable of mitigating frost formation through both passive and active approaches.

    The freezing process of liquid films and individual droplets on cold surfaces serves as a fundamental model for understanding the mechanisms of ice formation. Experimental studies have systematically explored this phenomenon using advanced imaging and measurement techniques. Early work by Anderson *et al.*[6] (1996) first observed the formation of pointed ice tips during droplet freezing. Subsequent analyses by Schultz *et al.*[7] established correlations between the tip geometry and interfacial curvature dynamics at moving phase boundaries. More recently, Hu *et al.*[8] revealed that freezing-induced expansion primarily occurs in the vertical direction rather than laterally. Chaudhary *et al.*[9] tracked temperature evolution using a combination of infrared cameras and thermocouples, while Zeng *et al.*[10] demonstrated how gravity

influences the droplet's initial morphology, thereby affecting its freezing patterns. Although these experimental observations have offered valuable empirical insights and validated the fundamental physical behaviors of ice formation processes, they remain largely limited to surface-level morphological characterization. Key internal characteristics, such as the three-phase contact line motion and the distributions of velocity and pressure, remain elusive due to current technological limitations. Furthermore, intrusive measurement techniques risk perturbing the system and inducing heterogeneous nucleation, thereby altering the natural crystallization kinetics under investigation. These dual constraints of restricted access to internal flow parameters and the risk of measurement-induced perturbations significantly hinder a comprehensive understanding of the multiphase dynamics involved in ice formation.

To overcome the limitations of experimental approaches, numerical methods have emerged as indispensable tools for probing the solidification of liquids in greater detail. Currently, various computational methods have been used, including the volume of fluid (VOF) method[11]-[14], the level-set method[15], [16], and the lattice Boltzmann method (LBM)[17]-[25]. For example, Lyu *et al*.[13] proposed a hybrid volume of fluid-immersed boundary (VOF-IB) method that takes into account volume expansion effects during droplet freezing, validating their model against existing experimental and theoretical results. Zhang *et al*.[14] coupled a VOF-based multiphase model with a solidification model in ANSYS FLUENT to simulate the freezing of water droplets, although their model neglected volume expansion effects. Based on the level-set method, Yan *et al*.[16] proposed a fully coupled computational model for thermal multiphase flow to simulate the melting and solidification processes of liquids. Meanwhile, Huang *et al*.[23] introduced a phase field model within the LBM framework to simulate unconfined droplet freezing on a plate. Xiong *et al*.[24] employed the pseudo-potential lattice Boltzmann method to systematically investigate the impact dynamics and solidification characteristics of droplets impinging on both smooth and rough cold substrates. Mohammadipour *et al*.[25] proposed an improved LBM for liquid solidification by modifying the convection term in the Cahn-Hilliard equation.

Despite these advancements, each of the aforementioned methods has inherent limitations. For example, the VOF method may suffer from numerical instability and convergence difficulties due to the strong coupling of multi-physics fields. Additionally, its interface reconstruction often struggles to accurately capture sharp or highly nonlinear solid-liquid interfaces, particularly during dynamic solidification processes. The level-set method typically requires frequent reinitialization, which can lead to mass loss near the interface. The LBM, while versatile, tends to exhibit limited numerical stability when dealing with multi-field coupling problems, especially under conditions involving large density ratios. Furthermore, because it relies on evolving discrete velocity distribution functions to update the flow field, the LBM incurs substantial memory overhead.

To overcome these drawbacks, a robust, accurate, and efficient numerical method is essential. The multiphase lattice Boltzmann flux solver (MLBFS), proposed by Wang et al.[26], offers a compelling alternative. As a finite volume-based solver, the MLBFS directly updates macroscopic flow variables at the cell centers. Unlike the traditional LBM, the MLBFS allows for the additional terms in the governing equations to be calculated as source terms and the physical boundary conditions to be applied directly through macroscopic quantities, avoiding the need for conversion of them into the forms of distribution functions. Furthermore, since MLBFS relies solely on equilibrium distribution functions related to conserved quantities, it significantly reduces memory consumption in the calculation as compared to conventional LBM. These advantages have led to the widespread application of MLBFS in simulating multiphase flow problems[27]-[35], particularly in scenarios involving large density and viscosity ratios, where it demonstrates excellent stability and mass conservation.

Building upon these unique strengths, this paper extends the MLBFS framework to simulate liquid solidification by developing an enthalpy-based unified lattice Boltzmann flux solver (EULBFS). In this method, the temperature field is governed by the enthalpy equation, the flow field by the Navier-Stokes equations, and the phase field by the Cahn-Hilliard (CH) equation. To ensure mass conservation in the process of liquid solidification, a volume expansion term is introduced into the continuity equation. Meanwhile, to ensure momentum conservation, a corresponding source term is incorporated into the momentum equation. All these macroscopic governing equations are discretized using the finite volume method (FVM), while the local solutions of the corresponding lattice Boltzmann equations (LBEs) are employed to calculate fluxes at the cell interface, ensuring that the solutions at the cell interface are consistent with the resolved macroscopic governing equations. The reliability and performance of the developed solver are validated through a series of benchmarks, involving the conductive freezing problem, the three-phase Stefan problem, the freezing of a liquid film in a two-dimensional container, the solidification of a static droplet on a cold surface, and the freezing of a droplet upon impact with a cold surface.

## 2. Enthalpy-Based Unified Lattice Boltzmann Flux Solver

### 2.1 Temperature field
#### 2.1.1 Enthalpy-based equation for liquid solidification

For the enthalpy-based solidification model, the corresponding macroscopic energy equation can be written as[23]:

$$\partial_t(\rho h) + \nabla \cdot (\rho h \mathbf{u}) = \nabla \cdot (\lambda \nabla T) + \rho h \nabla \cdot \mathbf{u} + Q_T \tag{1}$$

Here, $t$ is the time. $\rho$ and $\mathbf{u}$ are the fluid density and velocity vector, respectively. $h$, $T$, and $\lambda$ are the sensible enthalpy, temperature, and thermal conductivity. $Q_T$ is the heat source term caused by the absorption or release of latent heat, defined as:

$$Q_T = -\left[\partial_t(\rho \Delta H) + \nabla \cdot (\rho \mathbf{u} \Delta H)\right] \tag{2}$$

where $\Delta H$ is the latent enthalpy for phase change. For pure materials with uniform latent heat, the second term in Eq. (2) can be neglected. Introducing the liquid fraction $F_l = \Delta H / L$, where $L$ is the latent heat, the heat source term can be simplified to:

$$Q_T = -\partial_t(\rho L F_l) \tag{3}$$

Under the condition of incompressible flow, substituting Eq (3) into Eq (1) yields the total enthalpy-based energy equation:

$$\partial_t H + \nabla \cdot (C_p T \mathbf{u}) = \nabla \cdot \left( \frac{\lambda}{\rho} \nabla T \right) + C_p T \nabla \cdot \mathbf{u} \tag{4}$$

where $H$ is the total enthalpy, which can be expressed as $H = h + \Delta H = C_p T + LF_l$. $C_p$ is the specific heat at constant pressure. The liquid fraction and temperature can be determined from the total enthalpy $H$ by:

$$F_l = \begin{cases} 0 & H < H_s \\ \dfrac{H - H_s}{H_l - H_s} & H_s \leq H \leq H_l \\ 1 & H > H_l \end{cases} \tag{5}$$

$$T = \begin{cases} \dfrac{H}{C_p} & H < H_s \\ T_s + \dfrac{H - H_s}{H_l - H_s}(T_l - T_s) & H_s \leq H \leq H_l \\ T_l + \dfrac{H - H_l}{C_p} & H > H_l \end{cases} \tag{6}$$

where $H_s = C_{p,s} T_s$ is the total enthalpy at solidus temperature and $H_l = C_{p,l} T_l + L$ is the total enthalpy at liquid temperature. The thermophysical properties are determined based on the order parameter $C$ and solid fraction $F_s$:

$$\gamma = F_s \gamma_s + (1 - F_s) C \gamma_l + (1 - F_s)(1 - C) \gamma_g \tag{7}$$

where $\gamma$ denotes the density $\rho$, viscosity $\mu$, thermal conductivity $\lambda$, or heat capacity $C_p$. The subscripts $s$, $l$, and $g$ refer to solid, liquid, and gas phases, respectively. The solid fraction $F_s$ is calculated as $F_s = (1-F_l)$. The total enthalpy equation captures the phase change by coupling the temperature field with the phase fraction, thereby reducing the complexity of interface tracking.

**2.1.2 Governing equation for temperature field in EULBFS**

The standard lattice Boltzmann equation (LBE) for the total enthalpy distribution function $h_\alpha$ can be expressed as:

$$h_\alpha(\mathbf{x}+\mathbf{e}_\alpha\delta_t, t+\delta_t) - h_\alpha(\mathbf{x},t) = \frac{h_\alpha^{eq}(\mathbf{x},t) - h_\alpha(\mathbf{x},t)}{\tau_h}, \quad \alpha = 0-8 \tag{8}$$

where $h_\alpha$ and $h_\alpha^{eq}$ are the distribution function for the total enthalpy and its equilibrium state along the $\alpha$ direction, respectively. $\tau_h = 0.5 + \lambda/(\rho C_{p,ref} c_s^2 \delta_t)$ is the single relaxation parameter for the total enthalpy. $C_{p,ref}=0.5(C_{p,s}+C_{p,l})$ is the harmonic mean of specific heat. Here $C_{p,s}$ is the solid phase specific heat and $C_{p,l}$ is the liquid phase specific heat. $\delta_t$ is the streaming time step size. $\mathbf{e}_\alpha$ is the lattice speed. The equilibrium state $h_\alpha^{eq}$ used in this work is given by[36]:

$$h_\alpha^{eq} = \begin{cases} H - C_{p,ref}T + \omega_\alpha C_p T \left( \dfrac{C_{p,ref}}{C_p} - \dfrac{\mathbf{u}^2}{2c_s^2} \right), & \alpha = 0 \\ \omega_\alpha C_p T \left( \dfrac{C_{p,ref}}{C_p} + \dfrac{\mathbf{e}_\alpha \cdot \mathbf{u}}{c_s^2} + \dfrac{(\mathbf{e}_\alpha \cdot \mathbf{u})^2}{2c_s^4} - \dfrac{\mathbf{u}^2}{2c_s^2} \right), & \alpha = 1-8 \end{cases} \tag{9}$$

where $\omega_\alpha$ is the weight coefficient and $c_s$ is the sound speed. The discrete velocity distribution is shown in Fig. 1, with the velocity set defined as follows[37]:

$$\mathbf{e}_\alpha = \begin{cases} 0, & \alpha=0 \\ (\pm 1, 0), (0, \pm 1), & \alpha = 1-4 \\ (\pm 1, \pm 1), & \alpha = 5-8 \end{cases} \tag{10}$$

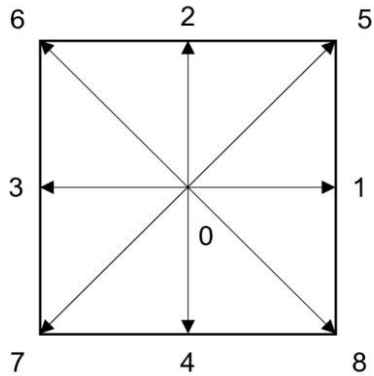

Fig 1. Diagram of the D2Q9 model.

This equilibrium distribution satisfies the following moment conditions:

$$\begin{cases} \sum_{\alpha} h_{\alpha}^{eq} = H \\ \sum_{\alpha} \mathbf{e}_{\alpha} h_{\alpha}^{eq} = C_p T \mathbf{u} \\ \sum_{\alpha} \mathbf{e}_{\alpha} \mathbf{e}_{\alpha} h_{\alpha}^{eq} = C_p T \mathbf{uu} + C_{p,ref} c_s^2 \mathbf{I} \end{cases} \quad (11)$$

Applying the multi-scale Chapman-Enskog expansion analysis to Eq. (8) gives the following equation (details in Appendix A):

$$\partial_t H + \nabla \cdot \left[ \sum_{\alpha} \mathbf{e}_{\alpha} \left( h_{\alpha}^{eq} + \left(1 - \frac{1}{2\tau_h}\right) h_{\alpha}^{neq} \right) \right] = 0 \quad (12)$$

where $h_{\alpha}^{neq}$ is the non-equilibrium distribution function for the total enthalpy, which can be calculated by $h_{\alpha}^{eq}$ at the cell interface and its surrounding points:

$$h_{\alpha}^{neq}(\mathbf{x},t) = -\tau_h \left[ h_{\alpha}^{eq}(\mathbf{x},t) - h_{\alpha}^{eq}(\mathbf{x} - \mathbf{e}_{\alpha}\delta_t, t - \delta_t) \right] \quad (13)$$

As shown in Appendix A, Eq (12) can be further rewritten as:

$$\partial_t H + \nabla \cdot (C_p T \mathbf{u}) = \nabla \cdot \left[ \frac{\lambda}{\rho} \nabla(T) \right] \quad (14)$$

It can be seen that the governing Eq. (14) recovered by the standard LBE is different from the target Eq. (4). In conventional LBM, this discrepancy is typically addressed by introducing an additional term into the LBE, which compromises the inherent simplicity of the method and may introduce additional errors.

To resolve this, the EULBFS directly incorporates the flux relationships derived from the standard LBE into the macroscopic governing Eq. (4) and treats the additional term as a source. Comparing Eq. (12) and Eq. (14), we have the relationship between the distribution function and the macroscopic fluxes (details in Appendix A):

$$C_p T \mathbf{u} - \frac{\lambda}{\rho} \nabla(T) = \sum_{\alpha} \mathbf{e}_{\alpha} \left( h_{\alpha}^{eq} + \left(1 - \frac{1}{2\tau_h}\right) h_{\alpha}^{neq} \right) \quad (15)$$

Introducing a new distribution function $h_{\alpha}^{*}$:

$$h_{\alpha}^{*} = h_{\alpha}^{eq} + \left(1 - \frac{1}{2\tau_h}\right) h_{\alpha}^{neq} \quad (16)$$

and substituting Eqs. (15) and (16) into Eq. (4), the macroscopic governing equation for total enthalpy can be rewritten as:

$$\partial_t H + \nabla \cdot \sum_\alpha \mathbf{e}_\alpha h_\alpha^* - C_p T \nabla \cdot \mathbf{u} = 0 \tag{17}$$

Note that in Eq. (16), the distribution function $h_\alpha^*$ is given by the standard LBE. It is clear that the macroscopic fluxes in the total enthalpy equation can be directly computed by the moment of the distribution function $h_\alpha^*$. Eq. (17) serves as the governing equation for the temperature field in EULBFS.

## 2.2 Flow field
### 2.2.1 Incompressible Navier-Stokes equations

For incompressible immiscible multiphase flows, the macroscopic governing equations without volume change can be written as[38]:

$$\frac{\partial p}{\partial t} + \rho c_s^2 \nabla \cdot \mathbf{u} = 0 \tag{18}$$

$$\frac{\partial \rho \mathbf{u}}{\partial t} + \nabla \cdot (\rho \mathbf{u}\mathbf{u}) = -\nabla p + \nabla \cdot \left[ \mu \left( \nabla \mathbf{u} + (\nabla \mathbf{u})^T \right) \right] + \mathbf{F}_{surf} + \mathbf{G} \tag{19}$$

where $p$ is the pressure, $\mathbf{F}_{surf}$ is the surface tension force, and $\mathbf{G}$ is the gravity.

During solidification, volume changes occur due to the density difference between the solid and liquid phases, leading to a breakdown of the assumption of constant volume at the solid-liquid interface. The velocity divergence equation must therefore be derived from the mass conservation principle:

$$D_t(M) = D_t(M_s + M_l) = 0 \tag{20}$$

where $M_s$ and $M_l$ represent the masses of the solid and liquid phases, respectively, and are given by:

$$\begin{cases} M_s = \int_V \rho_s F_s dV \\ M_l = \int_V \rho_l F_l dV \end{cases} \tag{21}$$

Substituting Eq. (21) into Eq. (20) yields:

$$D_t(M) = D_t \left[ \int_V (\rho_s F_s + \rho_l F_l) dV \right] = 0 \tag{22}$$

Applying the Reynolds transport theorem, the following equation can be obtained:

$$\int_{V(t)} \left[ \partial_t (\rho_s F_s) + \nabla \cdot (\rho_l F_l \mathbf{u}_l) - \partial_t (\rho_l F_s) \right] dV = 0 \qquad (23)$$

To ensure mass conservation over the control volume, the following relationship must hold:

$$\nabla \cdot (F_l \mathbf{u}_l) = \left(1 - \frac{\rho_s}{\rho_l}\right) \partial_t (F_s) \qquad (24)$$

Furthermore, since the total velocity has a relationship with the liquid velocity and the solid velocity as $\mathbf{u} = \mathbf{u}_l F_l + \mathbf{u}_s F_s$, assuming the velocity at the solid phase is zero, Eq. (24) can be simplified as:

$$\nabla \cdot \mathbf{u} = \left(1 - \frac{\rho_s}{\rho_l}\right) \partial_t (F_s) \qquad (25)$$

The right-hand side of Eq. (25) means the volume change associated with solidification, which is vital to keep the mass conservation. Thus, to ensure mass conservation in the solidification process, Eq. (18) should be modified as follows:

$$\frac{\partial p}{\partial t} + \rho c_s^2 \nabla \cdot \mathbf{u} = \rho c_s^2 \left(1 - \frac{\rho_s}{\rho_l}\right) \partial_t (F_s) \qquad (26)$$

Similarly, to conserve momentum during the solidification process, a body force $\mathbf{F}_{IBM} = -\rho \mathbf{u} \partial_t F_s$ is introduced, as suggested in Ref. [13]. Consequently, Eq. (19) is updated to the following form:

$$\frac{\partial \rho \mathbf{u}}{\partial t} + \nabla \cdot (\rho \mathbf{u} \mathbf{u}) = -\nabla p + \nabla \cdot \left[ \mu \left( \nabla \mathbf{u} + (\nabla \mathbf{u})^T \right) \right] + \mathbf{F}_{surf} + \mathbf{G} + \mathbf{F}_{IBM} \qquad (27)$$

### 2.2.3 Governing equation for flow field in EULBFS

The standard LBE for the flow field distribution function $f_\alpha$ can be expressed as:

$$f_\alpha (\mathbf{x} + \mathbf{e}_\alpha \delta_t, t + \delta_t) - f_\alpha (\mathbf{x}, t) = \frac{f_\alpha^{eq}(\mathbf{x}, t) - f_\alpha (\mathbf{x}, t)}{\tau_f}, \alpha = 0-8 \qquad (28)$$

where $f_\alpha$ and $f_\alpha^{eq}$ are the distribution function of the flow field and its equilibrium state along the $\alpha$ direction, respectively. $\tau_f = 0.5 + \mu/(\rho c_s^2 \delta_t)$ is the single relaxation parameter for the flow field. The equilibrium state $f_\alpha^{eq}$ is given by:

$$f_\alpha^{eq}(\mathbf{x},t) = \omega_\alpha \left[ p + \rho c_s^2 \left( \frac{(\mathbf{e}_\alpha \cdot \mathbf{u})}{c_s^2} + \frac{(\mathbf{e}_\alpha \cdot \mathbf{u})^2}{2c_s^4} - \frac{\mathbf{u}^2}{2c_s^2} \right) \right], \alpha = 0-8 \tag{29}$$

Using the Chapman-Enskog expansion analysis, the following macroscopic governing equations can be derived[26]:

$$\begin{cases} \dfrac{\partial p}{\partial t} + \nabla \cdot \left( \sum_\alpha \mathbf{e}_\alpha f_\alpha^{eq} \right) = 0 \\ \dfrac{\partial \rho c_s^2 \mathbf{u}}{\partial t} + \nabla \cdot \left[ \sum_\alpha \mathbf{e}_\alpha \mathbf{e}_\alpha \left[ f_\alpha^{eq} - \left(1 - \dfrac{1}{2\tau_f}\right) f_\alpha^{neq} \right] - \mu \Pi^e \right] = 0 \end{cases} \tag{30}$$

where

$$\begin{cases} f_\alpha^{neq}(\mathbf{x},t) = -\tau_f \left[ f_\alpha^{eq}(\mathbf{x},t) - f_\alpha^{eq}(\mathbf{x} - \mathbf{e}_\alpha \delta_t, t - \delta_t) \right] \\ \Pi^e = \dfrac{1}{\rho} \mathbf{u} \cdot \nabla (p - \rho c_s^2) \end{cases} \tag{31}$$

The relations between the macroscopic fluxes and the distribution function of the flow field are given by:

$$\begin{cases} \rho \mathbf{u} c_s^2 = \sum_\alpha \mathbf{e}_\alpha f_\alpha^{eq} \\ \rho \mathbf{u}\mathbf{u} + p\mathbf{I} - \mu(\nabla \mathbf{u} + (\nabla \mathbf{u})^T) = \dfrac{1}{c_s^2} \left[ \sum_\alpha \mathbf{e}_\alpha \mathbf{e}_\alpha \left[ f_\alpha^{eq} - \left(1 - \dfrac{1}{2\tau_f}\right) f_\alpha^{neq} \right] - \mu \Pi^e \right] \end{cases} \tag{32}$$

To eliminate the unwanted term of $\Pi^e$, Wang et al.[30] introduced a new distribution function at the surrounding points of the cell interface:

$$f_\alpha^{eq-m}(\mathbf{x} - \mathbf{e}_\alpha \delta_t, t - \delta_t) = \omega_\alpha \left[ p + \rho(\mathbf{x}, t - \delta_t) \Gamma_\alpha \right] \tag{33}$$

where

$$\Gamma_\alpha = c_s^2 \left[ 1 + \frac{\mathbf{e}_\alpha \cdot \mathbf{u}}{c_s^2} + \frac{(\mathbf{e}_\alpha \cdot \mathbf{u})^2}{2c_s^4} - \frac{\mathbf{u}^2}{2c_s^2} \right] \tag{34}$$

Using $f_\alpha^{eq-m}$, a new non-equilibrium term can be defined as:

$$f_\alpha^{neq-m}(\mathbf{x},t) = -\tau_f \left[ f_\alpha^{eq}(\mathbf{x},t) - f_\alpha^{eq-m}(\mathbf{x} - \mathbf{e}_\alpha \delta_t, t - \delta_t) \right] \tag{35}$$

The flux in Eq. (19) can then be calculated as:

$$\rho\mathbf{u}\mathbf{u}+p\mathbf{I}+\mu\left(\nabla\mathbf{u}+(\nabla\mathbf{u})^T\right)=\frac{1}{c_s^2}\left[\sum_\alpha \mathbf{e}_\alpha\mathbf{e}_\alpha\left[f_\alpha^{eq}-\left(1-\frac{1}{2\tau_f}\right)f_\alpha^{neq\_m}\right]\right] \quad (36)$$

Finally, by introducing a new distribution function $f_\alpha^*$:

$$f_\alpha^* = f_\alpha^{eq} - \left(1-\frac{1}{2\tau_f}\right)f_\alpha^{neq\_m} \quad (37)$$

The macroscopic governing equations of the flow field can ultimately be expressed as:

$$\begin{cases} \dfrac{\partial p}{\partial t}+\nabla\cdot\left(\sum_\alpha \mathbf{e}_\alpha f_\alpha^{eq}\right)-\mathbf{u}\cdot\nabla\rho c_s^2-\rho c_s^2\left(1-\dfrac{\rho_s}{\rho_l}\right)\partial_t(F_s)=0 \\ \dfrac{\partial \rho c_s^2 \mathbf{u}}{\partial t}+\nabla\cdot\left(\sum_\alpha \mathbf{e}_\alpha\mathbf{e}_\alpha f_\alpha^*\right)-c_s^2\left(\mathbf{F}_{surf}+\mathbf{G}+\mathbf{F}_{IBM}\right)=0 \end{cases} \quad (38)$$

The above set of equations constitutes the governing equations for the flow field in EULBFS, ensuring both mass and momentum conservation.

### 2.3 Phase field
#### 2.3.1 Cahn-Hilliard equation

The Cahn–Hilliard (CH) equation is adopted to track the evolution of the phase interface[39]. The corresponding governing equation can be expressed as:

$$\frac{\partial C}{\partial t}+\nabla\cdot(C\mathbf{u})=\nabla\cdot(M_c\nabla\mu_c)+C\nabla\cdot\mathbf{u} \quad (39)$$

where $C$ is the order parameter, $M_c$ is the mobility and $\mu_c$ is the chemical potential. It is important to note that the last term on the right-hand side of Eq. (39) must be included when accounting for volume change during the solidification process. By incorporating Eqs. (25) and (39), the CH equation with volume change can be written as:

$$\frac{\partial C}{\partial t}+\nabla\cdot(C\mathbf{u})=\nabla\cdot(M_c\nabla\mu_c)+C\left(1-\frac{\rho_s}{\rho_l}\right)\partial_t(F_s) \quad (40)$$

The chemical potential is defined as $\mu_c = 2\beta C(C-1)(2C-1)+\kappa\nabla^2 C$. Here, $\beta=12\sigma/\xi$ and $\kappa=3\xi\sigma/2$, where $\sigma$ is the interfacial tension coefficient and $\xi$ is the interface thickness.

## 2.3.2 Governing equation for phase field in EULBFS

Within the framework of LBM, Eq. (40) can be derived from the following LBE for the order parameter $C$:

$$g_\alpha(\mathbf{x}+\mathbf{e}_\alpha\delta_t, t+\delta_t) - g_\alpha(\mathbf{x},t) = \frac{g_\alpha^{eq}(\mathbf{x},t) - g_\alpha(\mathbf{x},t)}{\tau_g}, \alpha = 0-8 \quad (41)$$

where $g_\alpha$ and $g_\alpha^{eq}$ represent the distribution function of the order parameter and its equilibrium state, respectively. $\tau_g$ is the single relaxation parameter for the phase field, which is related to the diffusion parameter $Q_g$ and the mobility $M_c$ as $M_c = (\tau_g - 0.5)Q_g\delta_t$. The equilibrium state of the order parameter is given by:

$$g_\alpha^{eq}(\mathbf{x},t) = \begin{cases} C - \dfrac{\mu_c Q_g (1-\omega_0)}{c_s^2} & \alpha=0 \\ \dfrac{\omega_\alpha(\mu_c Q_g + C\mathbf{e}_\alpha \cdot \mathbf{u})}{c_s^2} & \alpha=1-8 \end{cases} \quad (42)$$

Using the Chapman-Enskog expansion analysis, the following macroscopic equation can be derived[39]:

$$\partial_t C + \nabla \cdot \sum_\alpha \mathbf{e}_\alpha \left[ g_\alpha^{eq} + \left(1 - \frac{1}{2\tau_g}\right) g_\alpha^{neq} \right] = 0 \quad (43)$$

where

$$g_\alpha^{neq}(\mathbf{x},t) = -\tau_g \left[ g_\alpha^{eq}(\mathbf{x},t) - g_\alpha^{eq}(\mathbf{x}-\mathbf{e}_\alpha\delta_t, t-\delta_t) \right] \quad (44)$$

By comparing Eqs. (39) and (43), the relationship between the macroscopic flux in the CH equation and the distribution function of the order parameter is established as:

$$C\mathbf{u} - M_c(\nabla\mu_c) = \sum_\alpha \mathbf{e}_\alpha \left[ g_\alpha^{eq} + \left(1 - \frac{1}{2\tau_g}\right) g_\alpha^{neq} \right] \quad (45)$$

By defining a new distribution function $g_\alpha^*$ as:

$$g_\alpha^* = \left[ g_\alpha^{eq} + \left(1 - \frac{1}{2\tau_g}\right) g_\alpha^{neq} \right] \quad (46)$$

and substituting Eqs. (45) and (46) into Eq. (40), the macroscopic governing equation for the phase field can be reformulated as:

$$\partial_t C + \nabla \cdot \sum_\alpha \mathbf{e}_\alpha g_\alpha^* - C\left(1 - \frac{\rho_s}{\rho_l}\right)\partial_t(F_s) = 0 \tag{47}$$

This equation represents the governing equation for the phase field in EULBFS.

### 2.4 Finite volume discretization of macroscopic governing equations

In EULBFS, the macroscopic governing equations for temperature, flow, and phase fields are solved using the cell-centered FVM. For clarity, Eqs. (17), (38) and (47) are rewritten in the following unified format:

$$\frac{\partial \mathbf{W}}{\partial t} + \nabla \cdot \mathbf{F} = \mathbf{F}_E \tag{48}$$

$$\mathbf{W} = \begin{pmatrix} H \\ p \\ \rho c_s^2 u \\ \rho c_s^2 v \\ C \end{pmatrix} \tag{49}$$

$$\mathbf{F} = \begin{pmatrix} \sum_\alpha \mathbf{e}_\alpha h_\alpha^* \\ \sum_\alpha \mathbf{e}_\alpha f_\alpha^{eq} \\ \sum_\alpha \mathbf{e}_\alpha e_{\alpha x} f_\alpha^* \\ \sum_\alpha \mathbf{e}_\alpha e_{\alpha y} f_\alpha^* \\ \sum_\alpha \mathbf{e}_\alpha g_\alpha^* \end{pmatrix} \tag{50}$$

$$\mathbf{F}_E = \begin{pmatrix} C_p T \nabla \cdot \mathbf{u} \\ \mathbf{u} \cdot \nabla \rho c_s^2 + \rho c_s^2\left(1 - \frac{\rho_s}{\rho_l}\right)\partial_t(F_s) \\ c_s^2\left(F_{surf,x} + F_{IBM,x}\right) \\ c_s^2\left(F_{surf,y} + G_y + F_{IBM,y}\right) \\ 0 \end{pmatrix} \tag{51}$$

Eq. (50) indicates that the fluxes of the macroscopic governing equations can be calculated by the moments of distribution functions at the cell interface, where the standard LBE is satisfied. This use of locally reconstructed LBE solutions at cell interfaces for calculating macroscopic fluxes is a key feature of EULBFS.

By integrating Eq. (48) over a control volume $i$, the following semi-discretized form of the macroscopic governing equations can be obtained:

$$\frac{d\mathbf{W}_i}{dt} = -\frac{1}{V_i} \sum_{j \in N(i)} \left( n_x \mathbf{F}_x + n_y \mathbf{F}_y \right)_{ij} S_{ij} + \mathbf{F}_{E,i} \tag{52}$$

with the terms are defined as:

$$\mathbf{F}_x = \begin{pmatrix} \sum_\alpha e_{\alpha x} h_\alpha^* \\ \sum_\alpha e_{\alpha x} f_\alpha^{eq} \\ \sum_\alpha e_{\alpha x} e_{\alpha x} f_\alpha^* \\ \sum_\alpha e_{\alpha x} e_{\alpha y} f_\alpha^* \\ \sum_\alpha e_{\alpha x} g_\alpha^* \end{pmatrix}, \quad \mathbf{F}_y = \begin{pmatrix} \sum_\alpha e_{\alpha y} h_\alpha^* \\ \sum_\alpha e_{\alpha y} f_\alpha^{eq} \\ \sum_\alpha e_{\alpha y} e_{\alpha x} f_\alpha^* \\ \sum_\alpha e_{\alpha y} e_{\alpha y} f_\alpha^* \\ \sum_\alpha e_{\alpha y} g_\alpha^* \end{pmatrix} \tag{53}$$

where $V_i$ is the volume of control cell $i$, $N(i)$ is the set of neighboring cells of cell $i$, and $S_{ij}$ is the area of the interface shared between cell $i$ and cell $j$. $\mathbf{n}_{ij}=(n_x, n_y)$ denotes the unit outer normal vector at the cell surface, pointing from cell $i$ to cell $j$. To solve Eq. (52), the fluxes at the cell interface $\mathbf{F}_x$ and $\mathbf{F}_y$, as well as the source term at the cell center $\mathbf{F}_E$, have to be determined first.

As shown in Eq. (53), the fluxes of $\mathbf{F}_x$ and $\mathbf{F}_y$ are the functions of $h_\alpha^*$, $f_\alpha^*$, and $g_\alpha^*$, which in turn depend on the equilibrium states at the cell interface and its surrounding points, according to Eqs. (16), (37) and (46). Thus, the equilibrium states at the relevant locations must be computed in advance. As illustrated in Fig. 2, $k_\alpha^{eq}$ represents the equilibrium states for temperature, flow, and phase fields, which are functions of the macroscopic flow variables. The primary task is thus to calculate these macroscopic variables at the corresponding locations.

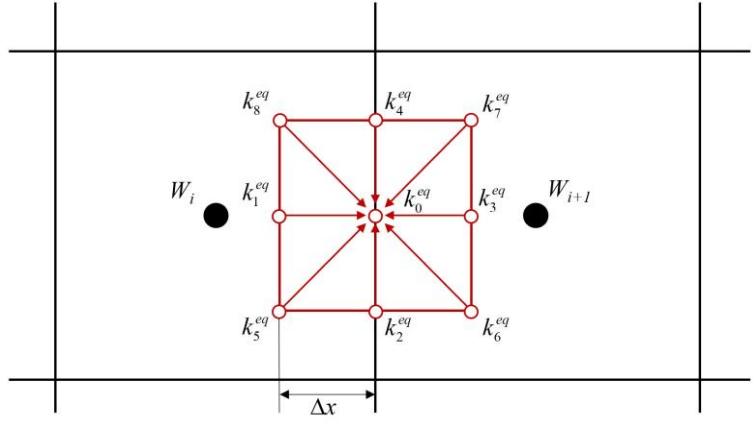

Fig 2. Local reconstruction of LBE solution at a cell interface.

In the cell-centered FVM, all macroscopic variables are defined at the cell centers. To obtain their values at the surrounding points of the cell interface, the following interpolation method is employed:

$$\psi(\mathbf{x}-\mathbf{e}_\alpha \delta_t, t-\delta_t) = \begin{cases} \psi(\mathbf{x}_i) + (\mathbf{x}-\mathbf{e}_\alpha \delta_t - \mathbf{x}_i) \cdot \nabla \psi(\mathbf{x}_i), & (\mathbf{x}-\mathbf{e}_\alpha \Delta t) \in W_i \\ \psi(\mathbf{x}_{i+1}) + (\mathbf{x}-\mathbf{e}_\alpha \delta_t - \mathbf{x}_{i+1}) \cdot \nabla \psi(\mathbf{x}_{i+1}), & (\mathbf{x}-\mathbf{e}_\alpha \Delta t) \in W_{i+1} \end{cases} \quad (54)$$

where $\mathbf{x}_i$, $\mathbf{x}_{i+1}$, and $\mathbf{x}$ represent the physical positions of two adjacent cell centers and their shared interface, respectively. $\psi$ denotes the physical variables such as $H$, $\mathbf{u}$, $p$, and $C$. Note that for the variables at the cell interface at the current time level, their values are taken as the average of the interpolated results from the left and right cells. After interpolation, the equilibrium states at the surrounding points of the cell interface $h_\alpha^{eq}(\mathbf{x}-\mathbf{e}_\alpha \delta_t, t-\delta_t)$, $f_\alpha^{eq}(\mathbf{x}-\mathbf{e}_\alpha \delta_t, t-\delta_t)$ and $g_\alpha^{eq}(\mathbf{x}-\mathbf{e}_\alpha \delta_t, t-\delta_t)$ can be obtained through Eqs. (9), (29) and (42), and the newly defined distribution function $f_\alpha^{eq-m}(\mathbf{x}-\mathbf{e}_\alpha \delta_t, t-\delta_t)$ can be calculated from Eq. (33).

According to the compatibility conditions, the collision terms on the right-hand side of Eqs. (8), (28) and (41) do not contribute to the calculation of conservation variables, and the following relations hold:

$$\begin{cases} \sum_\alpha h_\alpha^{neq}(\mathbf{x},t) = -\tau_h \sum_\alpha \left[ h_\alpha^{eq}(\mathbf{x},t) - h_\alpha^{eq}(\mathbf{x}-\mathbf{e}_\alpha \delta_t, t-\delta_t) \right] = 0 \\ \sum_\alpha f_\alpha^{neq}(\mathbf{x},t) = -\tau_f \sum_\alpha \left[ f_\alpha^{eq}(\mathbf{x},t) - f_\alpha^{eq}(\mathbf{x}-\mathbf{e}_\alpha \delta_t, t-\delta_t) \right] = 0 \\ \sum_\alpha \mathbf{e}_\alpha f_\alpha^{neq}(\mathbf{x},t) = -\tau_f \sum_\alpha \mathbf{e}_\alpha \left[ f_\alpha^{eq}(\mathbf{x},t) - f_\alpha^{eq}(\mathbf{x}-\mathbf{e}_\alpha \delta_t, t-\delta_t) \right] = 0 \\ \sum_\alpha g_\alpha^{neq}(\mathbf{x},t) = -\tau_g \sum_\alpha \left[ g_\alpha^{eq}(\mathbf{x},t) - g_\alpha^{eq}(\mathbf{x}-\mathbf{e}_\alpha \delta_t, t-\delta_t) \right] = 0 \end{cases}$$
(55)

Using these conditions, the macroscopic variables at the cell interface at the new time level can be computed from the surrounding equilibrium states:

$$\begin{cases} H(\mathbf{x},t) = \sum_\alpha h_\alpha^{eq}(\mathbf{x},t) = \sum_\alpha h_\alpha^{eq}(\mathbf{x}-\mathbf{e}_\alpha \delta_t, t-\delta_t) \\ p(\mathbf{x},t) = \sum_\alpha f_\alpha^{eq}(\mathbf{x},t) = \sum_\alpha f_\alpha^{eq}(\mathbf{x}-\mathbf{e}_\alpha \delta_t, t-\delta_t) \\ \rho(\mathbf{x},t)\mathbf{u}(\mathbf{x},t)c_s^2 = \sum_\alpha \mathbf{e}_\alpha f_\alpha^{eq}(\mathbf{x},t) = \sum_\alpha \mathbf{e}_\alpha f_\alpha^{eq}(\mathbf{x}-\mathbf{e}_\alpha \delta_t, t-\delta_t) \\ C(\mathbf{x},t) = \sum_\alpha g_\alpha^{eq}(\mathbf{x},t) = \sum_\alpha g_\alpha^{eq}(\mathbf{x}-\mathbf{e}_\alpha \delta_t, t-\delta_t) \end{cases}$$
(56)

Once the macroscopic variables at the cell interface at the new time level are determined, the corresponding equilibrium states $h_\alpha^{eq}(\mathbf{x},t)$, $f_\alpha^{eq}(\mathbf{x},t)$ and $g_\alpha^{eq}(\mathbf{x},t)$ can be calculated. Substituting these equilibrium states into Eqs. (16), (37) and (46), $h_\alpha^*$, $f_\alpha^*$ and $g_\alpha^*$ can be obtained. The macroscopic fluxes in Eq. (53) are then determined using $h_\alpha^*$, $f_\alpha^*$ and $g_\alpha^*$. In addition, the first and second terms in Eq. (51), $(C_p T \nabla \cdot \mathbf{u})$ and $\left[ \mathbf{u} \cdot \nabla \rho c_s^2 + \rho c_s^2 \left(1-\dfrac{\rho_s}{\rho_l}\right) \partial_t (F_s) \right]$, representing the source terms related to temperature and flow fields, can be directly calculated at the cell centers. Moreover, the surface tension can be evaluated as $\mathbf{F}_{surf} = -C\nabla \mu_c$.

Once both the macroscopic fluxes and the source terms are determined, Eq. (52) can be rewritten in the simplified residual form:

$$D_t \mathbf{W}_i = \mathbf{R}(\mathbf{W})$$
(57)

where **R(W)** encompasses the fluxes $\nabla \cdot \mathbf{F}$ and the source term $\mathbf{F}_E$. The above equation can be solved using the third-order Runge-Kutta method[40]:

$$\begin{cases} \mathbf{W}^{(1)} = \mathbf{W}^n + \Delta t \times \mathbf{R}(\mathbf{W}) \\ \mathbf{W}^{(2)} = \dfrac{3}{4}\mathbf{W}^n + \dfrac{1}{4}\mathbf{W}^{(1)} + \dfrac{1}{4}\Delta t \times \mathbf{R}(\mathbf{W}^{(1)}) \\ \mathbf{W}^{n+1} = \dfrac{1}{3}\mathbf{W}^n + \dfrac{2}{3}\mathbf{W}^{(2)} + \dfrac{2}{3}\Delta t \times \mathbf{R}(\mathbf{W}^{(2)}) \end{cases} \quad (58)$$

The time step size $\Delta t$ is determined by the Courant–Friedrichs–Lewy (CFL) condition. It should be noted that $\Delta t$ corresponds to the update time of the flow field physical quantities, whereas $\delta_t$ represents the streaming time step. The location $(\mathbf{x}-\mathbf{e}_\alpha \delta_t)$ should be within the appropriate range of the cells $W_i$ and $W_{i+1}$.

## 2.5 Computational procedures

The computational procedures for simulating the liquid solidification using the EULBFS can be summarized as follows:

(1) Specify the streaming time step $\delta_t$. Ensure that the surrounding points $(\mathbf{x}-\mathbf{e}_\alpha \delta_t)$ are within the appropriate range of the cell $W_i$ and $W_{i+1}$. Then, calculate the single-relaxation-time parameters $\tau_h$, $\tau_f$, and $\tau_g$.

(2) Interpolate the physical quantities $H$, $T$, $F_s$, $C_p$, $p$, $\mathbf{u}$, $C$, and $\mu_c$ at $(\mathbf{x}-\mathbf{e}_\alpha \delta_t)$ using Eq. (54). Then, evaluate the corresponding equilibrium states $h_\alpha^{eq}(\mathbf{x}-\mathbf{e}_\alpha \delta_t, t-\delta_t)$, $f_\alpha^{eq}(\mathbf{x}-\mathbf{e}_\alpha \delta_t, t-\delta_t)$, $f_\alpha^{eq\_m}(\mathbf{x}-\mathbf{e}_\alpha \delta_t, t-\delta_t)$ and $g_\alpha^{eq}(\mathbf{x}-\mathbf{e}_\alpha \delta_t, t-\delta_t)$ by Eqs. (9), (29), (33) and (42).

(3) Compute the macroscopic quantities at the center of the interface between two adjacent cells by Eq. (56). Then, calculate the equilibrium states $h_\alpha^{eq}(\mathbf{x},t)$, $f_\alpha^{eq}(\mathbf{x},t)$ and $g_\alpha^{eq}(\mathbf{x},t)$ at this location by Eqs. (9), (29) and (42).

(4) Evaluate the non-equilibrium terms $h_\alpha^{neq}$, $f_\alpha^{neq\_m}$, and $g_\alpha^{neq}$ by Eqs. (13), (35) and (44). Then, compute $h_\alpha^*$, $f_\alpha^*$ and $g_\alpha^*$ by Eqs. (16), (37) and (46).

(5) Calculate the fluxes **F** at the cell interfaces and the source terms $\mathbf{F}_E$ at the cell

centers. Then, solve the semi-discretized form of the macroscopic governing equations via the third-order Runge–Kutta method, as described in Eq. (58).

(6) Update the temperature field and liquid fraction by Eqs. (5) and (6). Subsequently, update the flow and phase fields accordingly to reflect the new state of the system.

(7) If the simulation time has reached the prescribed end time, terminate the simulation; otherwise, return to step (1) and proceed to the next time step.

## 3. Numerical Results and Discussions

In this section, the EULBFS is validated by various numerical tests, including the conductive freezing problem, the three-phase Stefan problem, the freezing of a liquid film in a two-dimensional container, the solidification of a static droplet on a cold surface, and the freezing of a droplet upon impact with a cold surface. The results obtained by the EULBFS are compared with the available analytical, numerical, and experimental results data to demonstrate its accuracy and reliability.

### 3.1 Conductive freezing problem

First, the conductive freezing of a pure substance is simulated to assess the accuracy of EULBFS in predicting temperature fields. As shown in Fig. 3, the computational domain is filled with liquid at a temperature $T_0$. $T_m$ is the melting temperature which is higher than $T_0$. The temperature at the left wall is maintained at $T_b$ ($T_b < T_m$), and the right wall is held at $T_0$ ($T_0 > T_m$). The velocity field is set to zero, and Dirichlet and periodic boundary conditions are applied in the $x$ and $y$ directions, respectively. The analytical solution for the temperature distribution is given by[23]:

$$T(X,t) = \begin{cases} T_b - \dfrac{(T_b - T_m)\operatorname{erf}\left(\dfrac{x}{2\sqrt{\alpha_s t}}\right)}{\operatorname{erf}(k)}, & 0 < X < X_i(t) \\ T_0 + \dfrac{(T_m - T_b)\operatorname{erf}\left(\dfrac{x}{2\sqrt{\alpha_l t}}\right)}{\operatorname{erf}\left(k\sqrt{\dfrac{\alpha_l}{\alpha_s}}\right)}, & X > X_i(t) \end{cases} \quad (59)$$

The position of the solid-liquid interface at time $t$ can be obtained by $X_i(t) = 2k\sqrt{\alpha_s t}$. Here, $\alpha$ is the thermal diffusivity, and $k$ is a constant determined by solving the following transcendental equation:

$$\frac{C_{p,s}(T_m - T_b)}{L\exp(k^2)\operatorname{erf}(k)} - \frac{C_{p,l}(T_0 - T_m)\sqrt{\frac{\alpha_l}{\alpha_s}}}{L\exp\left(\frac{k^2\alpha_s}{\alpha_l}\right)\operatorname{erf}\left(k\sqrt{\frac{\alpha_s}{\alpha_l}}\right)} = k\sqrt{\pi} \tag{60}$$

The physical parameters used in this case are: $\rho_l = \rho_s = 1$, $C_{p,s} = C_{p,l} = 1$, $\alpha_s = \alpha_l = 0.4$, $L=250$, $T_b=-1$, $T_0=1$, and $T_m=0$. Fig. 4 shows the comparison of the temperature distributions predicted by EULBFS and the analytical solution, The results exhibit excellent agreement, confirming the capability of the EULBFS in accurately capturing heat conduction and solidification in a pure substance.

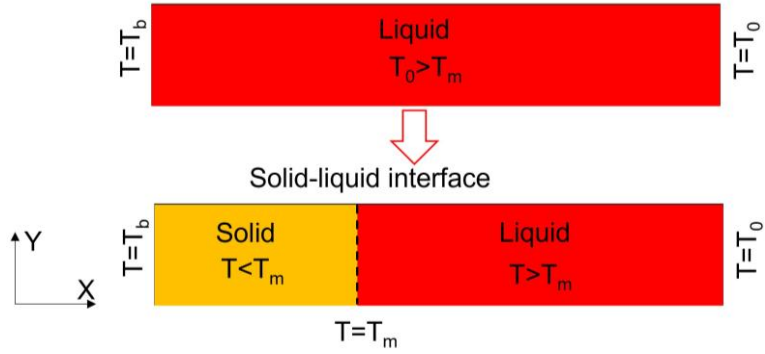

Fig 3. Configuration of the conductive freezing problem.

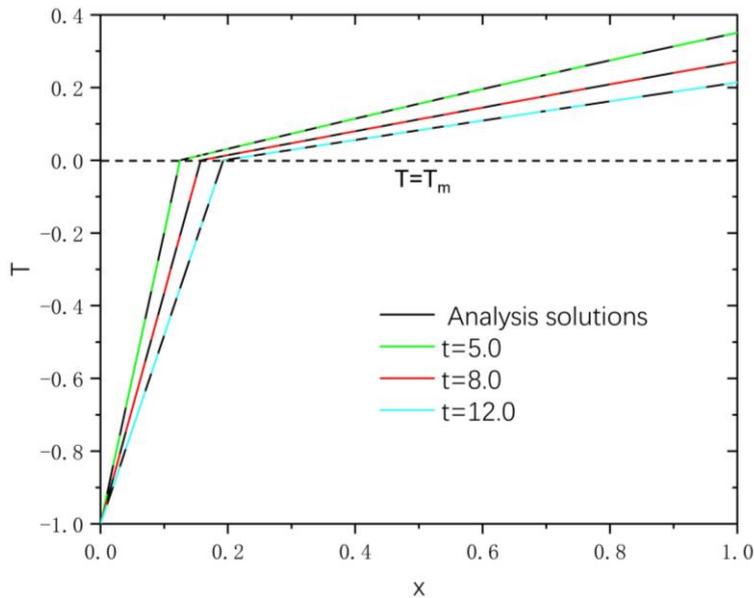

Fig 4. Comparison of temperature profiles obtained by EULBFS and the analytical solution.

**3.2 Three-phase Stefan problem**

To validate the ability of EULBFS to capture volume changes that happen during liquid solidification, a three-phase Stefan problem is simulated. The schematic diagram of the problem is shown in Fig. 5. Initially, the domain consists of a uniformly distributed liquid and gas region at temperature $T_0$. The liquid occupies the region $0<Y<H_0$, while the gas occupies $H_0<Y<H$. A constant temperature $T_w$ is imposed at the bottom wall, initiating the solidification of the liquid due to thermal conduction. Once the liquid solidification is completed, the maximum height $H_f$ of the solid can be obtained. According to the mass conservation, $H_f$ can be calculated as $H_f = (\rho_l/\rho_s)H_0$. The order parameter in the initial state is defined by:

$$C(X,Y) = 0.5\left[1+\tanh\frac{2(H_0-Y)}{\xi}\right] \tag{61}$$

where $\xi = 4$ is the thickness of the interface. To quantify the relative importance of sensible heat to latent heat, the Stefan number is introduced as $Ste = C_p(T_m - T_w)/L$.

In this case, the parameters are set as: $\rho_s:\rho_l = 0.8, 0.9, 1, 1.1, 1.2$, $T_0$=0.1, $T_w$=-2, $L$=250, $T_s=T_l=T_m$=0, $C_{p,s}:C_{p,l}$=1:1, $\lambda_s:\lambda_l = 1:1$, Ste=0.1, 0.15, 0.2, 0.25. For the temperature field, the adiabatic condition is applied to the left, right, and top walls, and a cold temperature $T_w$ is applied to the bottom wall. For the flow field, the no-slip condition is applied to the top and bottom walls, and the periodic boundary condition is applied to the left and right walls. For the phase field, the no-flux boundary condition is applied for all walls. The whole domain is discretized using a $100\times 400$ uniform grid. Fig. 6 shows the comparison of $H_f/H_0$ predicted by EULBFS with the analytical solution. The results show excellent agreement, confirming that EULBFS accurately captures the volume change during the liquid solidification and satisfies the mass conservation.

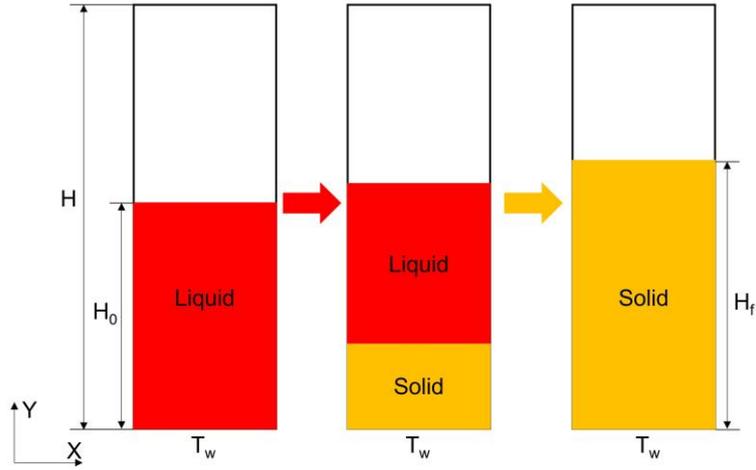

Fig 5. Schematic diagram of the three-phase Stefan problem.

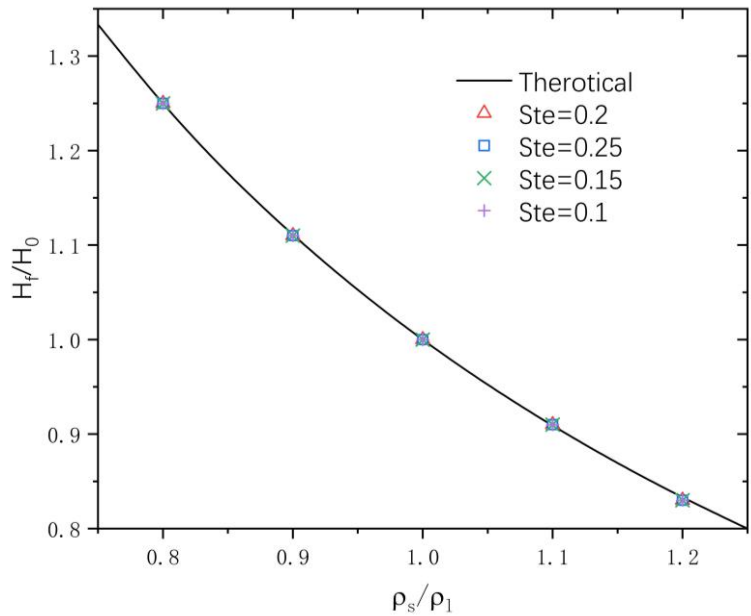

Fig 6. Comparison of normalized final height $H_f/H_0$ between the EULBFS results and the analytical solution for different density ratios $\rho_s/\rho_l$.

## 3.3 Freezing of a liquid film in a 2D container

To further validate the capability of the proposed solver in handling volume expansion during solidification, we simulate the freezing of a liquid film in a 2D container. Water is selected as the working liquid, and its physical properties are listed in Table 1. For the temperature field, the adiabatic condition is applied to the top wall, a symmetry condition is applied to the right wall, and a cold temperature ($T_w$=-20°C) condition is applied to the left and bottom walls. For the flow field, the outlet condition is applied to the top wall, a symmetry condition is applied to the right wall, and a no-slip condition is applied to the left and bottom walls. For the phase field, the no-flux boundary condition is applied for all walls. The initial temperature of the liquid is 20°C. The whole domain is discretized using a $200 \times 400$ uniform grid.

Table 1 Physical parameters for freezing of a liquid film in a 2D container.

| Properties | Water | |
| --- | --- | --- |
| | Liquid | Solid |
| Viscosity, $\mu(Pa\ s)$ | 0.0017 | - |
| Surface tension, $\sigma(N/m)$ | 0.076 | - |
| Density, $\rho(kg/m^3)$ | 1000 | 917 |
| Specific heat capacity, $C_p(J/(kg \cdot K))$ | 4210 | 2030 |
| Thermal conductivity, $\lambda(W/(m \cdot K))$ | 0.56 | 2.20 |
| Latent heat, $L(J/kg)$ | 334000 | - |
| Solidification temperature, $T_m(°C)$ | 0 | - |

Fig. 7 shows the solidification process for two different density ratios: (a) $\rho_s/\rho_l = 1$ and (b) $\rho_s/\rho_l = 0.917$. The solidification fronts are marked by the black dashed line. To quantitatively analyze the solidification process, the position evolutions of two special points, $h_w$ and $h_s$, measured from the left and bottom walls are introduced, as illustrated in Fig. 7(b). Fig. 8 presents the evolutions of $h_w$ and $h_s$. In the case of $\rho_s/\rho_l = 1$, there is no dilatation during the solidification process, and $h_w$ and $h_s$ remain the same. However, in the case of $\rho_s/\rho_l = 0.917$, the dilatation happens during the solidification process, pushing the liquid upward and causing the final solid shape to curve. This is due to the confinement imposed by the non-deformable left, bottom, and right walls, which restrict lateral expansion and force the dilatation in the vertical direction. Moreover, the case without the dilatation has a higher freezing ratio and a shorter frozen time. The results simulated by EULBFS are in good agreement with those reported by Lyu *et al.*[13]. The mass conservation test of the 2D liquid film is shown in Table 2, where $S_t$ and $S_n$ are the theoretical and numerical solid volumes, respectively. The relative error between $S_t$ and $S_n$ is calculated by $Err=|S_n-S_t|/S_t$. Both cases demonstrate excellent mass conservation properties.

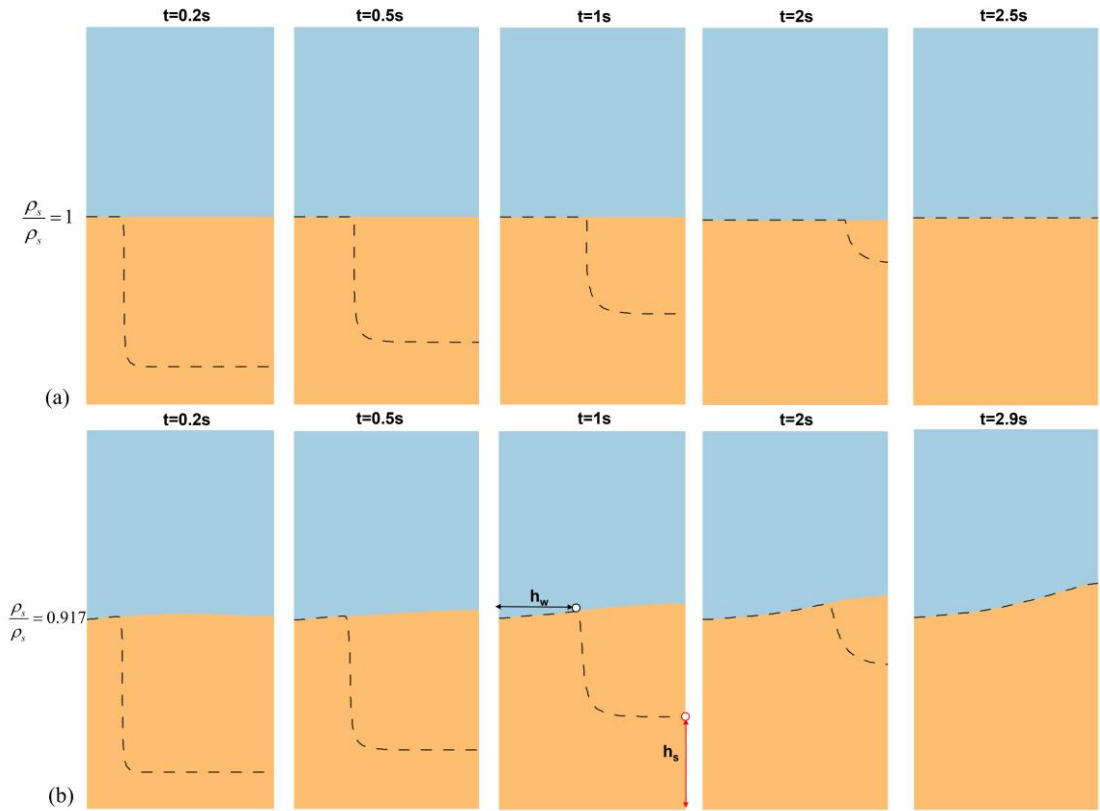

Fig.7 Freezing of a liquid film cooled from left and bottom walls.

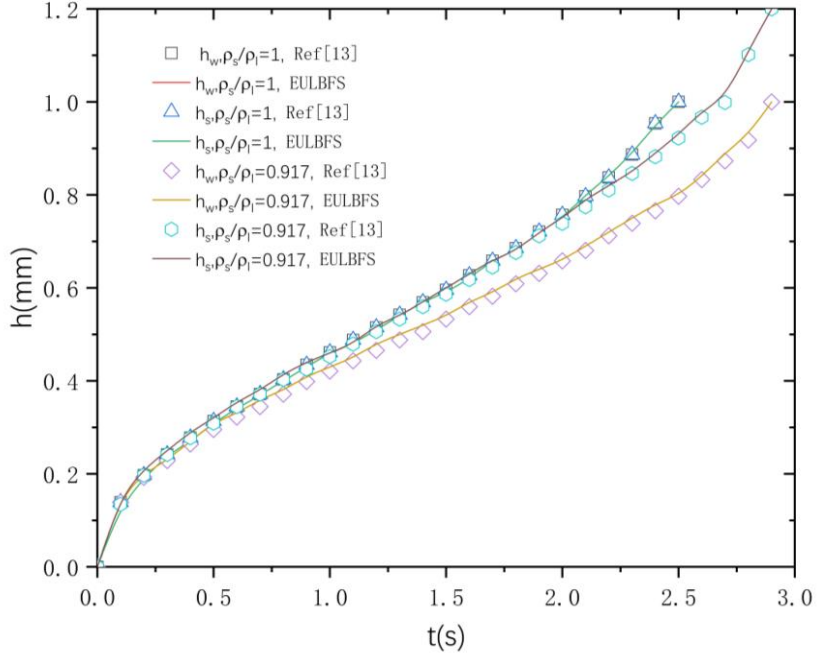

Fig.8 Position evolutions of $h_w$ and $h_s$ for cases with different density ratios $\rho_s/\rho_l = 1$ and $\rho_s/\rho_l = 0.917$.

Table 2 Mass conservation test for cases with different density ratios $\rho_s/\rho_l = 1$ and $\rho_s/\rho_l = 0.917$.

| $\rho_s/\rho_l$ | $S_t \left(mm^2\right)$ | $S_n \left(mm^2\right)$ | $Err = \left|S_n - S_t\right|/S_t$ |
| --- | --- | --- | --- |
| 1 | 1.0000 | 1.0000 | $1 \times 10^{-15}$% |
| 0.917 | 1.0907 | 1.0905 | 0.18% |

### 3.4 Solidification of a static droplet on a cold surface

In this section, we investigate the solidification of a static droplet on a cold surface to further validate the effectiveness of the EULBFS in capturing phase change dynamics. Specifically, the solidifications of both water and hexadecane droplets are considered. Due to the extremely short duration of the nucleation and recalescence stages relative to the overall solidification process, these early stages are not considered in the simulation. Instead, we focus on the subsequent solidification process, assuming heat balance and taking the pre-nucleation state as the initial condition. The physical properties of water and hexadecane droplets adopted in the simulation are listed in Table 3. The initial ambient temperature is set to 20°C. For the water droplet, the bottom wall is maintained at $T_w$=-10°C, while for the hexadecane droplet, $T_w$=15°C. The initial diameters of the water and hexadecane droplets are 2.84mm and 2.67mm, respectively. For the temperature field, the adiabatic condition is applied to the left, right, and top walls, and a cold temperature condition $T_w$ is applied to the bottom wall. For the flow field, the no-slip condition is applied to all walls. For the phase field, the no-flux boundary condition is applied for all walls. The whole domain is discretized using a 400×200 uniform grid. The order parameter in the initial state is given by:

$$C(X,Y) = 0.5\left[1 + \tanh\frac{2\left(R - (X-X_0)^2 - (Y-Y_0)^2\right)}{\xi}\right] \quad (62)$$

Table 3 Physical parameters of water and hexadecane droplets.

| Properties | Water | | Hexadecane | |
|---|---|---|---|---|
| | Liquid | Solid | Liquid | Solid |
| Viscosity, $\mu(Pa\ s)$ | 0.003 | - | 0.003 | - |
| Surface tension, $\sigma(N/m)$ | 0.076 | - | 0.028 | - |
| Density, $\rho(kg/m^3)$ | 999 | 917 | 774 | 833 |
| Specific heat capacity, $C_p(J/(kg\cdot K))$ | 4220 | 2100 | 2310 | 1800 |
| Thermal conductivity, $\lambda(W/(m\cdot K))$ | 0.581 | 2.16 | 0.15 | 0.15 |
| Latent heat, $L(J/kg)$ | 333400 | - | 230000 | - |
| Solidification temperature, $T_m(°C)$ | 0 | - | 18 | - |

Fig. 9 shows a comparison of droplet profiles and solidification interfaces between the simulations and experiments at different time steps. The black line marks the solid-liquid interface. Upon the onset of solidification, the interface progresses upward for both water and hexadecane droplets. In experimental images[41], the red line means the height of solidification at the interface, where the interface appears relatively flat due to optical obstruction by opaque ice. In contrast, the simulation results reveal a concave interface, aligning with the observations reported by Marin *et al*[42]. During the water droplet's solidification process, the interface expands outward and eventually forms a conical tip due to the lower density of ice compared to water and the influence of surface tension. In contrast, the hexadecane droplet forms a flatter interface that contracts inward toward the end of solidification. These findings confirm that the EULBFS can effectively capture solidification characteristics for fluids with varying density ratios.

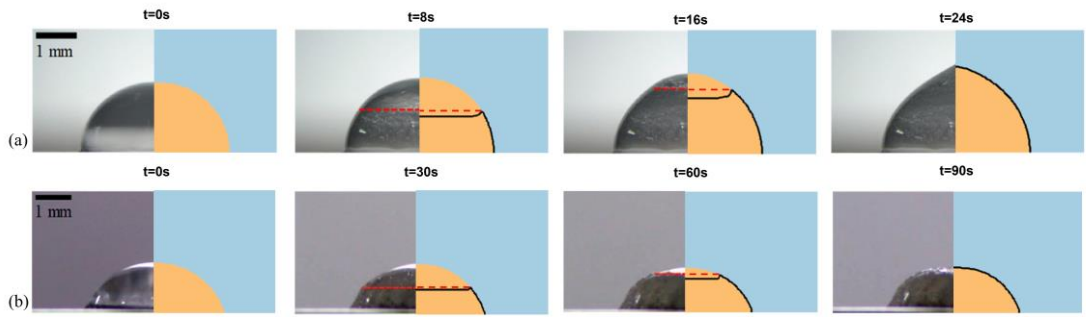

Fig.9 Solidification evolution of (a) water and (b) hexadecane droplets.

Fig. 10 displays the evolutions of the radius ($R_l$) and height ($H_l$) of the solidification interface of (a) water and (b) hexadecane droplets. The results obtained by EULBFS are compared with those from LBM[41] and experiment[41]. Minor deviations are observed between simulations and experiments due to the use of the diffuse interface[43]. To assess accuracy more quantitatively, the solidification times of water and hexadecane droplets are listed in Table 4. The EULBFS results exhibit a smaller deviation from experimental data than those from the LBM[41], indicating improved accuracy.

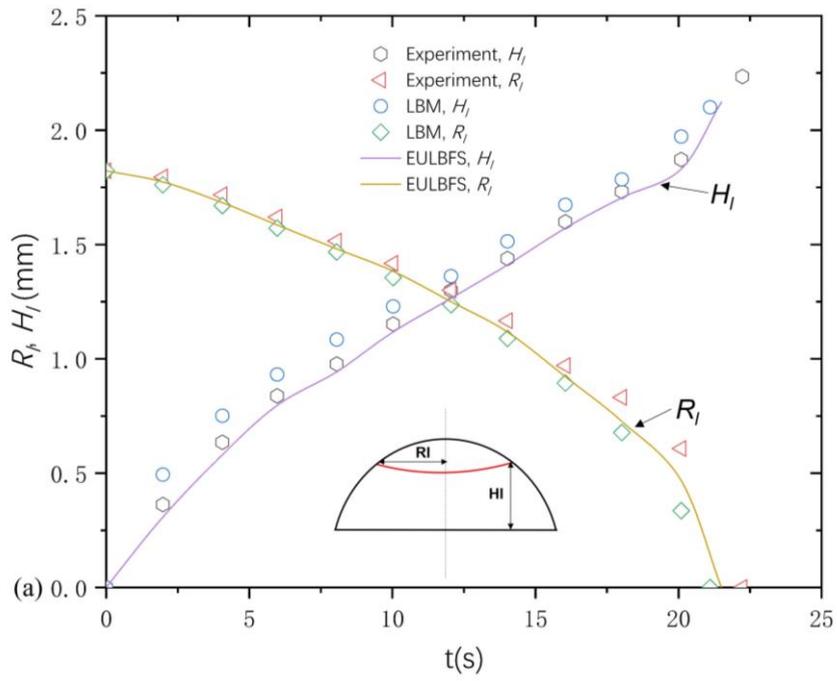

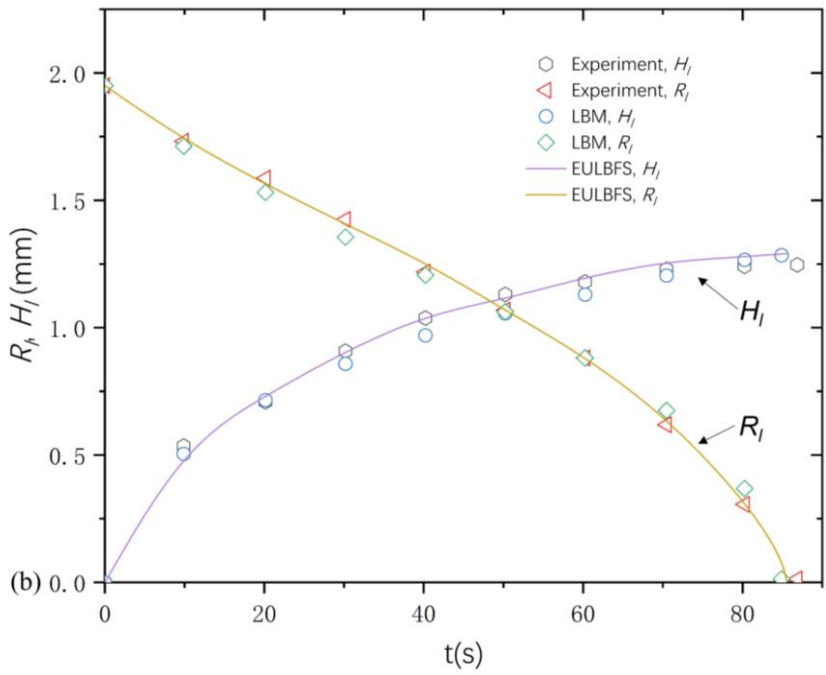

Fig.10 Comparison of solidification interface radius and height evolution among EULBFS, LBM, and experimental data for (a) water and (b) hexadecane droplets.

Table 4 Comparison of solidification time from experiment, LBM, and EULBFS.

|  | Solidification times(s) | |  | Solidification times(s) | |
| --- | --- | --- | --- | --- | --- |
|  | Water | Hexadecane |  | Water | Hexadecane |
| Experiments[41] | 22.2 | 86.6 | Experiments[41] | 22.2 | 86.6 |
| LBM[41] | 21.1 | 84.9 | EULBFS | 21.5 | 85.5 |
| Deviation | 5.05% | 2.00% | Deviation | 3.15% | 1.27% |

The final frozen shapes of droplets with various contact angles $\theta$ and density ratios are shown in Figs. 11 and 12. Droplets with larger contact angles exhibit smaller base radii and greater heights, which reduce the contact area with the cold surface and increase the thermal resistance. These factors decrease the latent release and prolong the solidification time. Table 5 shows the volume changes from the final to the initial stages. The maximum deviation is only 0.064%, confirming excellent mass conservation of the EULBFS across various contact angles and density ratios. Fig. 13 shows the final frozen shape of five different density ratios $\rho_s/\rho_l$ at $\theta=120°$. It can be seen that when lower solid-liquid density ratios ($\rho_s/\rho_l>1$) cause vertical expansion, while higher ratios ($\rho_s/\rho_l<1$) result in radial contraction. Overall, the expansion and contraction of the droplet mainly occur in the vertical direction, with minor changes observed radially. These trends align well with the previous experimental results[44]. All the results in this section show that the proposed EULBFS can accurately and effectively simulate the solidification dynamics of static droplets, including volume expansion effects, shape evolution, and mass conservation.

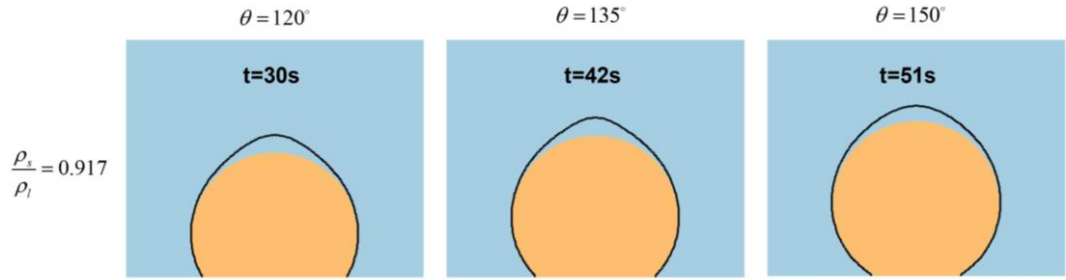

Fig.11 Final frozen shape of the water droplet at different contact angles.

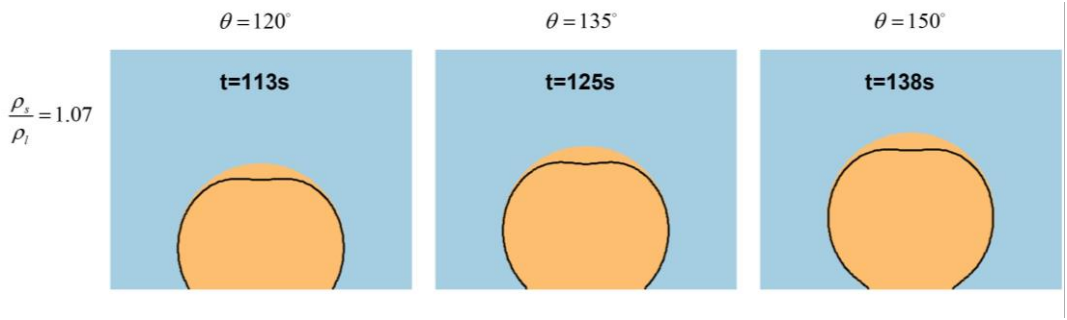

Fig.12 Final frozen shape of the hexadecane droplet at different contact angles.

Table 5 Volume change between initial and final stages.

|  | $\rho_s/\rho_l = 0.917$ | | | $\rho_s/\rho_l = 1.07$ | | |
| --- | --- | --- | --- | --- | --- | --- |
|  | $S_t$ | $S_n$ | Err | $S_t$ | $S_n$ | Err |
| $\theta = 120°$ | 1.0905 | 1.0907 | 0.018% | 0.9346 | 0.935 | 0.042% |
| $\theta = 135°$ | 1.0905 | 1.0901 | 0.036% | 0.9346 | 0.9342 | 0.042% |
| $\theta = 150°$ | 1.0905 | 1.0908 | 0.027% | 0.9346 | 0.934 | 0.064% |

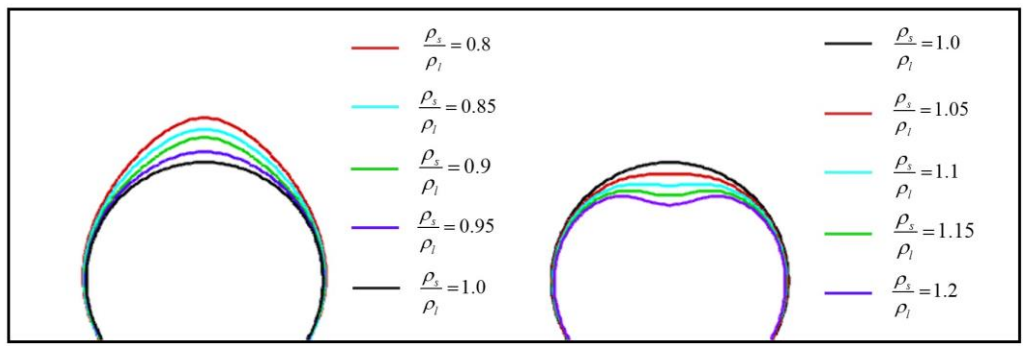

Fig 13. Influent of density ratio $\rho_s/\rho_l$ on the final frozen droplet shape.

## 3.5 Freezing of a droplet upon impact with a cold surface

To further validate the present model in dynamic conditions involving solidification, we investigate the freezing behavior of a droplet impacting a cold surface. Since the interface of the droplet evolves during the impact process, to ensure that the surface tension only acts on the non-frozen part, the expression of the surface tension is modified as $\mathbf{F}_{surf} = -F_l C \nabla \mu_c$ [23]. Fig. 14 shows the initial setup for this problem. For the temperature field, the adiabatic condition is applied to the left, right, and top walls, while a fixed cold temperature $T_w$ is applied to the bottom wall. For the flow field, the no-slip condition is applied for all walls. For the phase field, the no-flux boundary condition is applied. The whole domain is discretized using a $400 \times 200$ grid.

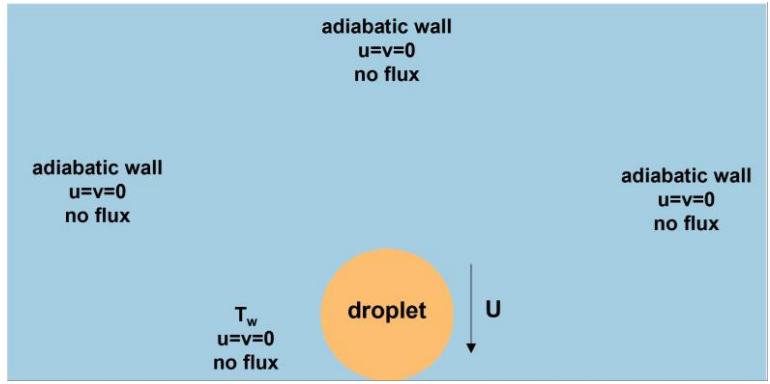

Fig. 14 Initial configuration of a droplet impacting a cold surface.

Fig. 15 shows the morphological evolution of the droplet during the spreading process, as simulated by the EULBFS, and compares it with the experimental observations by (a) Zhang $et\ al.$[45] and (b) Yao $et\ al.$[46]. The black line denotes the solidification interface. The parameters for the two cases are as follows:

Case (a): The initial temperature of the water droplet is -5°C, the temperature of the bottom wall is $T_w$=-30°C, the diameter of the water droplet is 2.84mm, and the impacting velocity is U=0.7m/s. The advancing contact angle is $\theta_{adv}=100°$, the receding contact angle is $\theta_r=39°$, and the equilibrium contact angle $\theta_{eq}=82°$.

Case (b): The initial temperature of the water droplet is 25°C, the temperature of the bottom wall is $T_w$=-30°C, the diameter of the water droplet is 2.62mm, and the impacting velocity is U=0.44m/s. The advancing contact angle is $\theta_{adv}=105°$, the receding contact angle $\theta_r = 70°$, and the equilibrium contact angle $\theta_{eq} = 70°$.

As can be seen from Fig. 15, the contact area increases upon impact. As the droplet reaches its maximum spread and begins to shrink, the top of the droplet rebounds. However, due to the formation of an ice layer at the base, the droplet is pinned and prevented from detaching from the cold surface. To quantitatively evaluate the spreading behavior, the spreading factor $D/D_0$ is defined, where D is the instantaneous spreading diameter and $D_0$ is the initial droplet diameter. Fig. 16 compares the temporal evolution of the spreading factor obtained from experiment and simulation. Table 6 lists the maximum spreading factor from both sources. The maximum deviation is under 4%, confirming the accuracy of EULBFS in solving droplet impact with solidification.

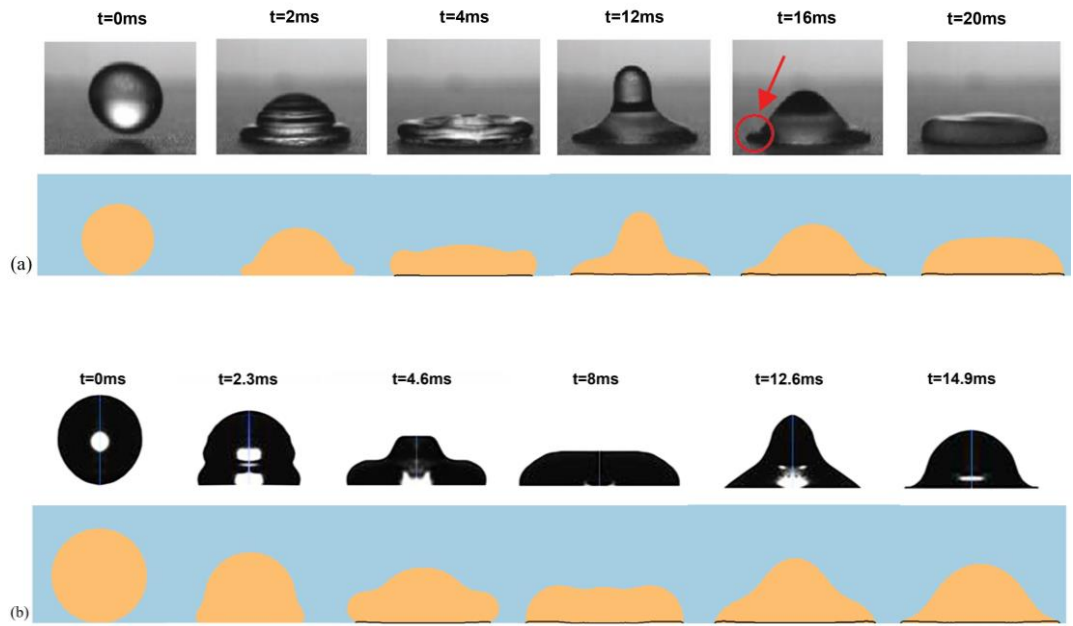

Fig. 15 Comparison of droplet shape evolution during impact between experiment and EULBFS simulation.

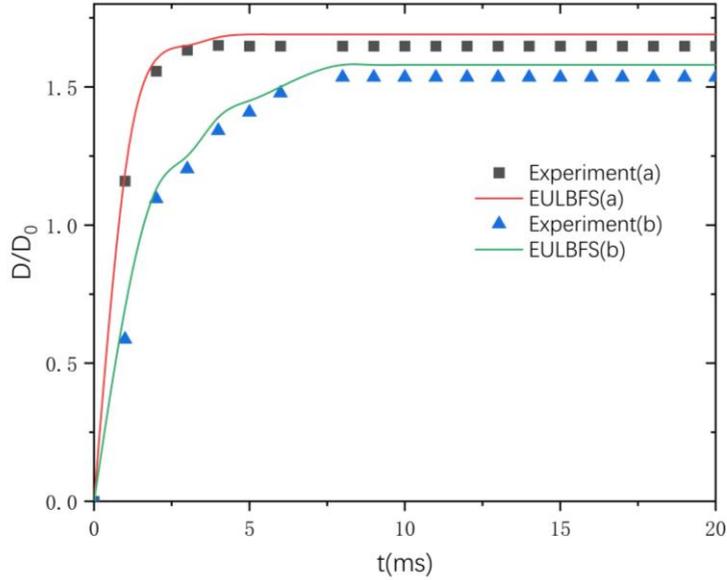

Fig. 16 Comparison of spreading factor evolution between experiment and EULBFS simulation.

Table 6 Comparison of maximum spreading factor between experiment and EULBFS simulation.

|          | Experiment | Simulation | Err   |
|----------|------------|------------|-------|
| Case(a)  | 1.65       | 1.7        | 3.03% |
| Case(b)  | 1.55       | 1.61       | 3.87% |

Fig. 17 shows the evolution of droplet impact on a supercooled plate at temperatures ranging from $T_w$=-20°C to $T_w$=-35°C, using parameters from the case (b). The corresponding spreading factors are plotted in Fig. 18. Initially, the spreading diameters are nearly identical across all temperatures. However, as the bottom wall temperature decreases, the earlier onset of solidification results in reduced maximum spreading. These findings are consistent with the experimental observations reported by Yao et al.[46]. Additionally, Table 7 lists the ratio of the droplet bounce height $H_b$ to the initial droplet diameter $D_0$ at t=15ms. From this table, it can be observed that the lower temperatures lead to greater rebound heights after maximum spreading, due to the more rapid formation of an ice base.

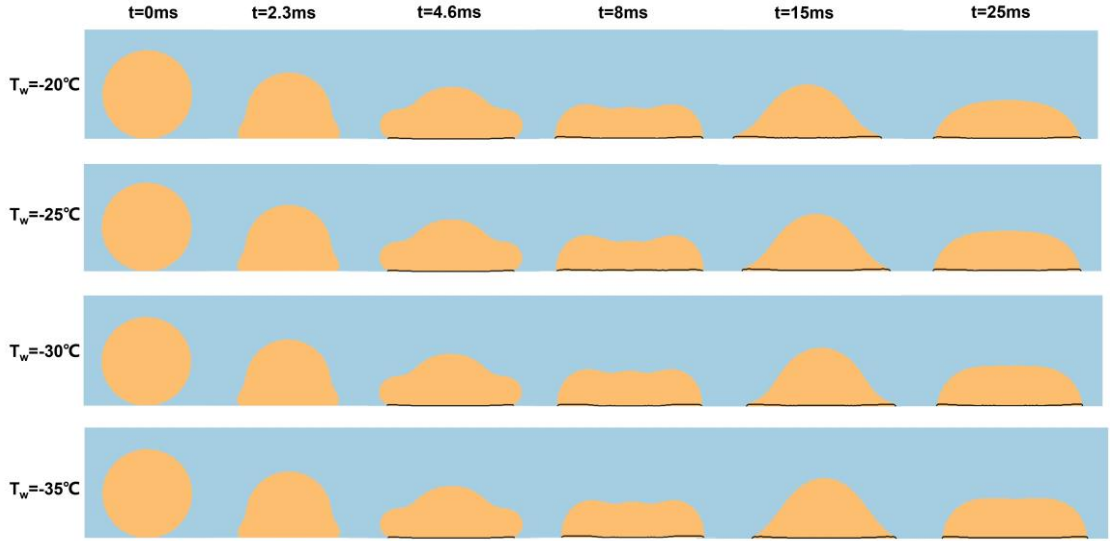

Fig. 17 Evolution of droplet impact on a supercooled plate at different bottom wall temperatures $T_w$.

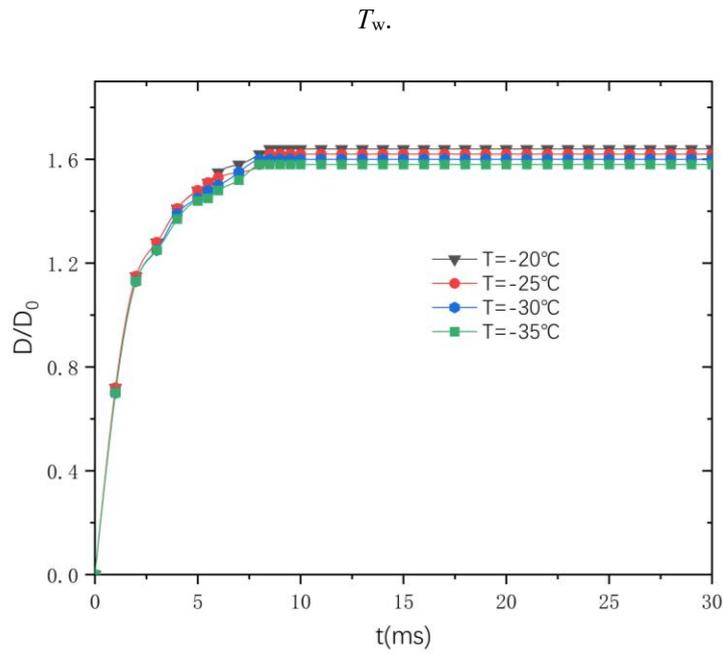

Fig. 18 Temporal variation of spreading factor for different bottom wall temperatures $T_w$.

Table 7 Comparison of $H_b/D_0$ at different $T_w$ at t=15ms.

|  | $T_w$=-20°C | $T_w$=-25°C | $T_w$=-30°C | $T_w$=-35°C |
|---|---|---|---|---|
| $H_b/D_0$ | 0.618 | 0.642 | 0.666 | 0.683 |

## 4. Conclusion

In this paper, we present an enthalpy-based uniform lattice Boltzmann flux solver (EULBFS) to effectively simulate incompressible liquid solidification. In this solver, the total enthalpy equation, NS equations, and CH equation are discretized using FVM. By locally reconstructing solutions to the LBEs corresponding to these macroscopic governing equations at the cell interfaces, the macroscopic numerical fluxes can be calculated efficiently and accurately. To account for the volume expansion or shrinkage caused by the density differences between the liquid and solid phases, an additional term is introduced into the continuity and momentum equations. This enhancement ensures a more realistic representation of the phase transition process.

The proposed method offers several advantages: (1) The total enthalpy equation directly describes the phase change through the coupling of the temperature field and phase fraction, which reduces the computational complexity of interface capturing; (2) The solver retains the advantages of standard LBM in the flux reconstruction process; (3) It provides a more straightforward implementation of external forces and boundary conditions. The EULBFS has been validated through a series of benchmark tests, including the conductive freezing problem, the three-phase Stefan problem, the freezing of a liquid film in a two-dimensional container, the solidification of a static droplet on a cold surface, and the freezing of a droplet upon impact with a cold surface. In all cases, the simulation results show good agreement with theoretical predictions, experimental observations, and previous numerical studies. These validations confirm the accuracy and robustness of EULBFS in capturing the liquid solidification process. Meanwhile, the method demonstrates excellent mass conservation properties, further supporting its reliability in practical applications.

## Appendix A. Chapman-Enskog expansion analysis

The Taylor series expansion of Eq. (8) can be expressed as:

$$\left(\partial_t + \mathbf{e}_\alpha \cdot \nabla\right) h_\alpha + \frac{1}{2} \delta_t \left(\partial_t + \mathbf{e}_\alpha \cdot \nabla\right)^2 h_\alpha + o\left(\delta_t^2\right) = -\frac{1}{\tau_h \delta_t} \left(h_\alpha - h_\alpha^{eq}\right) \qquad \text{(A-1)}$$

By introducing the multi-scale expansion:

$$\begin{cases} \partial_t = \varepsilon^1 \partial_{t1} + \varepsilon^2 \partial_{t2} \\ h_\alpha = h_\alpha^0 + \varepsilon^1 h_\alpha^1 + \varepsilon^2 h_\alpha^2 \\ \nabla_x = \varepsilon^1 \nabla_{x1} \end{cases} \quad (A\text{-}2)$$

where $\varepsilon$ is a small parameter proportional to the Knudsen number. Substituting Eq. (A-2) into (A-1), the equation can be decomposed into a hierarchy of equations at successive orders of $\varepsilon$:

$$\begin{cases} o(\varepsilon^0): h_\alpha^{(0)} = h_\alpha^{eq} \\ o(\varepsilon^1): \partial_{t_1} h_\alpha^{(0)} + \mathbf{e}_\alpha \cdot \nabla_{x1} h_\alpha^{(0)} = -\dfrac{1}{\tau_h \delta_t} h_\alpha^{(1)} \\ o(\varepsilon^2): \partial_{t2} h_\alpha^{(0)} + \partial_{t1}\left(1 - \dfrac{1}{2\tau_h}\right) h_\alpha^{(1)} + \mathbf{e}_\alpha \cdot \nabla_{x1}\left(1 - \dfrac{1}{2\tau_h}\right) h_\alpha^{(1)} = -\dfrac{1}{\tau_h \delta_t} h_\alpha^{(2)} \end{cases} \quad (A\text{-}3)$$

Utilizing Eq. (11) and $\sum_\alpha h_\alpha^{(n)} = 0$, Eq. (A-3) can be rewritten as:

$$\begin{cases} \varepsilon^1: \partial_{t_1} H + \nabla_{x1} \sum_\alpha \mathbf{e}_\alpha h_\alpha^{(0)} = 0 \\ \varepsilon^2: \partial_{t2} H + \nabla_{x1} \cdot \left[\left(1 - \dfrac{1}{2\tau_h}\right) \sum_\alpha \mathbf{e}_\alpha h_\alpha^{(1)}\right] = 0 \end{cases} \quad (A\text{-}4)$$

Combining the resultant formulations on the $t_0$ and $t_1$ time scales and defining $h_\alpha^{(1)} = h_\alpha^{neq}$, the following equation can be obtained:

$$\partial_t H + \nabla \cdot \left[\sum_\alpha \mathbf{e}_\alpha \left(h_\alpha^{eq} + \left(1 - \dfrac{1}{2\tau_h}\right) h_\alpha^{neq}\right)\right] = 0 \quad (A\text{-}5)$$

To derive the relationship between the distribution function and the macroscopic fluxes, substituting Eq. (11) into Eq. (A-4) yields:

$$\varepsilon^1: \partial_{t1} H + \nabla_{x1} \cdot (C_p T \mathbf{u}) = 0 \quad (A\text{-}6)$$

$$\varepsilon^2: \partial_{t2} H + \nabla_{x1} \cdot \left[\left(1 - \dfrac{1}{2\tau_h}\right) \sum_\alpha \mathbf{e}_\alpha h_\alpha^{(1)}\right] = 0 \quad (A\text{-}7)$$

According to Eqs. (11) and (A-3), we can deduce:

$$-\dfrac{1}{\tau_h \delta_t} \sum_\alpha \mathbf{e}_\alpha h_\alpha^{(1)} = \nabla_{x1} \cdot (C_{p,ref} T c_s^2 \mathbf{I}) + \partial_{t1}(C_p T \mathbf{u}) + \nabla_{x1} \cdot (C_p T \mathbf{u}\mathbf{u}) \quad (A\text{-}8)$$

Substituting Eq. (A-8) into (A-7) and neglecting the higher order terms $\partial_{t1}(C_p T\mathbf{u})$ and $\nabla_{x1} \cdot (C_p T\mathbf{uu})$[36], Eq. (A-7) can be simplified as:

$$\partial_{t2} H = \nabla_{x1} \cdot \left[ c_s^2 (\tau_h - 0.5) \delta_t \nabla_{x1} (C_{p,ref} T) \right] \quad \text{(A-9)}$$

It should be noted that $C_{p,ref}$ keeps unvaried over the entire space. Thus, combining Eqs. (A-6) and (A-9), the following equation can be obtained:

$$\partial_t H + \nabla \cdot (C_p T\mathbf{u}) = \nabla \cdot \left[ c_s^2 (\tau_h - 0.5) \delta_t C_{p,ref} \nabla (T) \right] \quad \text{(A-10)}$$

Moreover, the relationship between $\tau_h$ and $\lambda$ can be expressed as:

$$\tau_h = \frac{\lambda}{\rho c_s^2 \delta_t C_{p,ref}} + 0.5 \quad \text{(A-11)}$$

Substituting Eq. (A-11) into (A-10), Eq. (A-10) can be further simplified as:

$$\partial_t H + \nabla \cdot \left[ C_p T\mathbf{u} - \frac{\lambda}{\rho} \nabla (T) \right] = 0 \quad \text{(A-12)}$$

By comparing Eq. (A-5) and Eq. (A-12), the following relationship can be obtained:

$$C_p T\mathbf{u} - \frac{\lambda}{\rho} \nabla (T) = \sum_\alpha \mathbf{e}_\alpha \left( h_\alpha^{eq} + \left(1 - \frac{1}{2\tau_h}\right) h_\alpha^{neq} \right) \quad \text{(A-13)}$$

**Acknowledgments**

The research is supported by the National Natural Science Foundation of China (12202191, 92271103) and the State Key Laboratory of Mechanics and Control for Aerospace Structures (Nanjing University of Aeronautics and Astronautics) (MCAS-S-0324G03).